# EMPIRICAL STUDY OF SUSTAINING THE ACTUALIZED VALUE PROPOSITIONS OF IMPLEMENTED E-GOVERNMENT PROJECTS IN SUB-SAHARAN AFRICA

Yusuf Ephraim Chidama, American University of Nigeria Yola, yusuf.chidama@aun.edu.ng

Chidi Gerard Ononiwu, American University of Nigeria Yola, chidi.ononiwu@aun.edu.ng

**Abstract:** Governments in sub-Saharan Africa have implemented e-Government projects. Actualizing the value propositions and sustaining such values are becoming problematic. Some scanty studies on the value propositions of implemented e-Government projects did not consider actualization of the values. Besides, such studies lack theoretical underpinnings, the identification, and measure of what constitutes actualized values. Neither did they capture what mechanisms could sustain the actualized values nor the contextual conditions enabling its sustainability. Consequently, using a concept-centric systematic review, we identified the value proposition of such implemented projects. By drawing from theories of affordance actualization, realist evaluation (RE) theory, self-determination theory, and sustainability framework for e-Government success. We conducted a RE of the implemented e-Government projects in Rwanda using RE as a methodology in three phases. In phase one, we developed the initial program theory (IPT), in phase two, we used contingent valuation as a quantitative approach and realist interview as qualitative method to validate the IPT. Lastly, in the third phase, we synthesized the results of the two investigative case studies to develop the actualized values sustainability framework. Such framework encapsulates, the actualized value propositions, mechanisms and enabling conditions in interactions to sustain the value propositions discovered in the e-Government investigative contexts.

## 1. INTRODUCTION

There is an increasing scarce resource at the disposal of governments in emerging economies (Heeks, 2002; Heeks, 2017), and citizens' expectations for quality service delivery is continuously rising. Government plays pivotal role in delivery of efficient, effective and prompt services (Rose, Persson, Heeager, et al., 2015). With the pervasive and emergent role that information and communication technologies (ICTs) play in disruptive quality service delivery; governments in emerging economies are adopting ICTs projects in public sectors in the form of e-Government (Lessa & Tsegaye, 2019). Thus, robust and quality e-Government value propositions that improve service delivery to citizens are increasing in sub-Saharan Africa (Nkohkwo & Islam, 2013). Value propositions in this study identify how public organizations by use of e-Government fulfils citizens' needs across different roles (e.g., class, gender, ethnic, demography and business concerns). Examples are the Rwandan e-Government initiative known as Irembo, (meaning "access" in local language) and various e-Government initiatives in Nigeria such as the Immigration e-passport service. But sustaining them are becoming problematic (Klischewski & Lessa, 2013). Besides, value propositions are perceived to be a means of accommodating all dimensions of e-government's performance to prove its relevance to stakeholders. Such value propositions are also anchored on the capability of e-Government projects to actualize the different needs of stakeholders and deliver services they value most. Thereby legitimizing and sustaining the implementation of such projects (Twizeyimana & Andersson, 2019). Sustaining e-Government projects here is leveraging the existing structures, processes and resources associated with the implemented e-Government projects to continue to deliver the envisaged value propositions to the citizens (Pang et al., 2014). This is done over time, in a given context for the benefits of the stakeholders, which include citizens and donor agencies and to build local capacities (Pluye et al., 2004). Consequently, sustainability of such projects takes the shape of strengthening the system to constitute a resilient ICT4D project. Considering the presence of constrains in the developing economies' context of the sub-Saharan Africa where the system is operationally situated. Such constraints include cultural and political challenges as well as huge financial constraints. They also include widespread corruption, poor infrastructure, high inequalities, fragile democracies and information poverty within the context the innovation is embedded (Heeks, 2003, 2017). Besides, a large number of e-Government projects in developing countries have failed to enhance governance or underutilized by the host countries (Twizeyimana & Andersson, 2019). Despite huge resources committed with expectations dampened (Heeks, 2002; Heeks, 2017; Heeks & Bailur, 2007). Thus, sustainability of e-Government projects could be considered as resilience, where both concepts refer to the state of a system or feature over time. While focusing on the persistence of that system under normal operating conditions and in response to constraints, setbacks and changes (Ahern, 2013; Anderies et al., 2013; Pang et al., 2014).

On one hand, value propositions of e-Government projects can be created by the host countries to assume the service provision segment of e-Government value propositions (Grimsley & Meehan, 2007; Hui & Hayllar, 2010; Omar et al., 2011; Twizeyimana & Andersson, 2019). While on the other hand actualization of such created value propositions assumed the service consumption segment of the implemented e-Government projects of which it has proven difficult to sustain by the citizens (Verkijika & De Wet, 2018). Consequently, actualizing values propositions is the motivations and capabilities of the citizens to exploit, appropriate, and use the already implemented e-Government projects to access government services. This is done in an efficient and effective manner to enhance citizens' participation in governance, engage in civic responsibilities as well their general well-being.

Some scanty attempts by literature to uncover the sustainability of the actualized value propositions of the implemented e-Governments has meet stiff oppositions that hinder the advancement of research in this area (Bannister & Connolly, 2014; Heeks & Bailur, 2007). Such oppositions include: (1) lack of theoretical underpinnings (Bannister & Connolly, 2014; Heeks & Bailur, 2007), (2) the identification and measure of what really constitute the actualized value propositions by the citizens (3) what mechanisms could be in place to sustain the actualized values and lastly, under what contextual conditions could such sustainability be possible. Thus, to overcome these oppositions, we deem it fit to conduct an evidence-based theory driven realist evaluation of such projects (Mukumbang et al., 2016; Pawson & Tilley, 1997). Evidence-based here suggests that the study obtains empirical evidence on the relative operational effectiveness of the implemented e-Government projects. This should be across various stakeholders and what the technology holds for each stakeholder in their respective contexts of use. To do this, we pose the following research question: *How can the actualized value propositions of implemented e-Government projects in sub-Saharan Africa be sustained and under what conditions, and why?"*

To answer the research question, we draw from multiple theories of affordance actualization (Strong et al., 2014), realist evaluation theory (RE) (Pawson & Tilley, 1997), Self-determination theory (SDT) (Deci & Ryan, 1985), and Sustainability framework for e-Government success (Lessa et al., 2015). This is done in the Rwanda's implemented e-Government projects as the investigative contexts. This is because no single theory could be used to delineate the actualized values, its sustenance and in what contextual conditions. We got to know this from the initial empirical evidences collected when matched with the tentative theoretical concepts of the aforementioned theories in an abductive manner (Ononiwu et al., 2018). Each theory represents a version of the phenomenon in question (Ononiwu et al., 2018). Thus, to "develop deeper levels of explanation and understanding" (McEvoy & Richards, 2006, p. 69), we need such theories to afford a more compelling accounts of the actualized value propositions of implemented e-Government projects, its sustainability and under what conditions.

The remainder of this paper is structured as follows: Section 2 provides the systematic literature review adopted for the study. Section 3 presents the theoretical foundation. In section 4, we discuss the realist case study research methodology, while Section 5 holds the findings. The discussion is in Section 6, while Section 7 holds the theoretical and practical contribution of the study. We conclude our study and highlight the future research endeavor in Section 8.

## 2. LITERATURE REVIEW

To identify the value propositions, we conducted a concept-centric systematic literature review (Cram et al., 2017; Okoli, 2015; Schryen et al., 2017; Vom Brocke et al., 2015; Webster & Watson, 2002) followed by thematic coding analysis as advocated by IS researchers (Aksulu & Wade, 2010; Bandara et al., 2015; Inuwa & Ononiwu, 2020; Roberts et al., 2012). We conducted a keyword search in titles, abstracts, and references of researches archived in PUBMED, Wiley online, Emerald, Springer, Elsevier. Also, Taylor and Francis, (AISeL) and the latest version of e-Government Reference Library (EGRL) version 15.5. The keywords we used were; *"e-Government value proposition, public value of e-Government, e-Government evaluation, ICT enabled public administration, sub-Saharan Africa, Affordance actualization"*. Our initial search turned up 14,921 articles. After applying our inclusion criteria, which include: (a) published peer-reviewed articles, (b) articles written in English language, (c) articles published from 2005-2019, (d) theoretical articles or articles that tested and validated models and frameworks, (e) conceptually rich articles. The exclusion criteria are: (a) articles not written in English language, (b) articles published before 2005, (c) a-theoretical papers that are conceptually weak, and (d) research in progress as well as working papers. We were left with 136 articles; 61 duplicates were removed leaving 75 articles. After snowballing the articles, 6 more papers were added, having a total of 81 articles, 5 articles were not found and 76 articles were found to be relevant to the study. This leads to the emergence of our value propositions of implemented e-Government projects as shown in Table 1.

| Value Propositions | Authors |
|---|---|
| Quality Service Delivery | (Agbabiaka, 2018; Alruwaie et al., 2012; Ashaye & Irani, 2019; Bai, 2013; Bonina & Cordella, 2009; Castelnovo & Simonetta, 2008; Cook & Harrison, 2014; Deng et al., 2018; Flak et al., 2009; Golubeva & Gilenko, 2018; Golubeva et al., 2019; Grimsley & Meehan, 2007; 2008; Grimsley et al., 2006; Gupta & Suri, 2017; Hellang & Flak, 2012; Karkin & Janssen, 2014; Karunasena & Deng, 2010; Karunasena et al., 2011; Omar et al., 2011; Pang et al., 2014; Persson et al., 2017; Rose, Persson, & Heeager, 2015; Rose, Persson, Heeager, et al., 2015; Scott et al., 2016; Sundberg, 2019; Twizeyimana & Andersson, 2019) |
| Effective Public Organizations | (Agbabiaka, 2018; Alruwaie et al., 2012; Ashaye & Irani, 2019; Avdic & Lambrinos, 2015; Bai, 2013; Bonina & Cordella, 2009; Castelnovo, 2013; Chu et al., 2017; Cook & Harrison, 2014; Deng et al., 2018; Golubeva et al., 2019; Gupta & Suri, 2017; Harrison et al., 2012; Hellang & Flak, 2012; Jussila et al., 2017; Karunasena & Deng, 2011; Karunasena et al., 2011; Persson et al., 2017; Rose, Persson, & Heeager, 2015; Rose, Persson, Heeager, et al., 2015; Savoldelli et al., 2013; Scott et al., 2016; Srivastava, 2011; Tsohou et al., 2013; Twizeyimana & Andersson, 2019) |

| Social Value and General Wellbeing | (Agbabiaka, 2018; Alruwaie et al., 2012; Ashaye & Irani, 2019; Chircu, 2008; Golubeva & Gilenko, 2018; Golubeva et al., 2019; Harrison et al., 2012; Hellang & Flak, 2012; Hu et al., 2019; Hussein, 2018; Karunasena & Deng, 2012; Raus et al., 2009; Rose et al., 2018; Savoldelli et al., 2013; Scott et al., 2016; Sigwejo & Pather, 2016; Twizeyimana & Andersson, 2019; Uppström & Lönn, 2017) |
|---|---|
| Open Government and Democratic Value | (Agbabiaka, 2018; Ashaye & Irani, 2019; Avdic & Lambrinos, 2015; Bai, 2013; Chu et al., 2017; Cook & Harrison, 2014; Flak et al., 2009; Golubeva & Gilenko, 2018; Golubeva et al., 2019; Grimsley & Meehan, 2007; 2008; Grimsley et al., 2006; Harrison et al., 2012; Hellang & Flak, 2012; Hu et al., 2019; Hussein, 2018; Karunasena & Deng, 2012; Karunasena et al., 2011; Omar et al., 2011; Pang et al., 2014; Persson et al., 2017; Rose et al., 2018; Rose, Persson, & Heeager, 2015; Rose, Persson, Heeager, et al., 2015; Savoldelli et al., 2013; Scott et al., 2016; Subbiah & Ibrahim, 2011; Sundberg, 2019; Twizeyimana & Andersson, 2019; Wihlborg et al., 2017) |

**Table 1 Value Propositions and the Corresponding Authors**

We now briefly discussed the value propositions of the implemented e-Government projects gleaned from literature and categorized into themes. Such themes include (a) Quality service delivery, which reflects avenues for generating value propositions through delivery of better services (Bannister & Connolly, 2014; Castelnovo, 2013). (b) Effective public organizations as it delineates the e-Government capacity to improve efficiency of public organizations through cost cutting, achieving synergy and synchronizing public organizations (Heeks & Bailur, 2007). Efficiency in public organizations ensures that there is a reduction in duplicate tasks, effective prompt delivery of services and easy access to services (Pang et al., 2014). (c) Open government and democratic values. Within our literature review, we realized that IS scholars seek to refocus attention to a broader array of values. Most especially those concerning open government (OG) and democratic value (Deng et al., 2018; Harrison et al., 2012). The foundation of OG idea is the optimism over that which can be achieved through the use of e-Government open data initiative. The essence of OG is to avail information and decision-making processes of government accessible to the public for scrutiny and input (Deng et al., 2018). In so doing, citizen's social and political engagements are facilitated through e-participation in developing and executing government policies. (d) Finally, social value and wellbeing: IS scholars always advocate that the expectations of e-Government go beyond mere citizen satisfaction, but they should encompass a desire for much broader social outcomes (Grimsley & Meehan, 2008; Karunasena & Deng, 2012). Such social outcomes include social inclusion, community development, well-being and sustainability (Harrison et al., 2012). Besides, general wellbeing that are encapsulated in the form of quality of health care, threshold standards of education, and access to civil and criminal justice should not be left out (Castelnovo, 2013).

## 3.  THEORETICAL FOUNDATION

As stated earlier, the theories adopted for this research are theory of affordance actualization (Strong et al., 2014), realist evaluation theory (RE) (Pawson & Tilley, 1997), Self-determination theory (SDT) (Deci & Ryan, 1985) and Sustainability framework for e-Government success (Lessa et al., 2015). Affordance actualization theory (Strong et al., 2014) grounded in realist evaluation theory (Pawson & Tilley, 1997) guided us in investigating the process through which e-Government affordances were actualized to produce value propositions as outcomes. Thus, the integration of affordance actualization and realist evaluation theory enable us a five-fold understanding of the affordance actualization process: affordances, context, mechanisms, actualization actions and outcomes theory (Pawson & Tilley, 1997; Strong et al., 2014). Besides, actualization actions reveal the human agencies' (citizens) choice to use the technology, their motivations and capabilities to exploit, appropriate and use the affordances offered by the implemented e-Government projects. Consequently, the study introduced the Self-determination theory (SDT) (Deci & Ryan, 1985; Ryan & Deci, 2002). SDT is an eclectic theory of human motivation that examines how individuals interact with their environment; in this case, the environment is the implemented e-Government projects (Deci & Ryan, 1985; Ryan & Deci, 2002). SDT aided us in specifying the nature of individual competence and performance within the experiences of choices that determines their actions. Such determinants are autonomy, competency and relatedness (Deci & Ryan, 1985; Ryan & Deci, 2002). Autonomy is viewed as a need to feel free and possess self-directed cognizance in the environment, which signifies one's sense of control and agency (Ryan & Deci, 2002). Competency is a feel of being effective, broadly, a feeling of being competent with task, activities, and engagements (Ryan & Deci, 2002). While relatedness is outlined as a feeling of being included and affiliated with others. We used the sustainability framework for e-Government success (Lessa et al., 2015) to reveal the generative mechanisms to sustain the value propositions as outcomes from the affordance actualization process.

The sustainability framework for e-Government success (Lessa et al., 2015) provides rich concepts that characterized sustaining e-Government project. This is in terms of innovation continuation considered as mechanisms. Such mechanism concepts include: (a) developing a sense of national ownership (b) continuously meet available resources (c) independence from donor agencies/external assistance (d) continuous monitoring and evaluation (e) institutionalizing e-Government projects with local context (f) strong political support and leadership (g) availability of institutional, administrative and coordinated capacity (h) meeting stakeholders needs (Lessa et al., 2015).

We operationalized the sustainability of the implemented e-Government project into five dimensions. Such dimensions include: (a) endurance of the identified actualized values (Slaghuis et al., 2011). (b) persistence of routinization or institutionalization of the initial e-Government projects (Akaka et al., 2017). (c) continued adaptation within the context the innovation is embedded (Buchanan et al., 2006). (d) evolutionary growth through configuration (Fleiszer et al., 2015), and (e) maintenance and appropriation with local capacities over time (Heeks, 2005). Furthermore, the program theory of RE originally designed by Pawson et al. (2005) and modified by Mukumbang et al. (2016) was deployed as the methodology to uncover the actualized value propositions, what mechanisms could be in place to sustain the actualized values and under what contextual conditions could such sustainability be possible. .

## 4. RESEARCH METHODOLOGY

Being a RE approach, this study is anchored on the six stages iterative process of RE in a multiple case study environment. These stages are: (1) development of a preliminary program theory; (2) search strategy and literature search; (3) study selection and appraisal; (4) data extraction; 5. data analysis and synthesis; and (6) program theory reformulation and recommendations. However, following Mukumbang et al. (2016) the six stages are encapsulated in three phases and adopted as the methodology to answer our research questions as shown in Figure 1

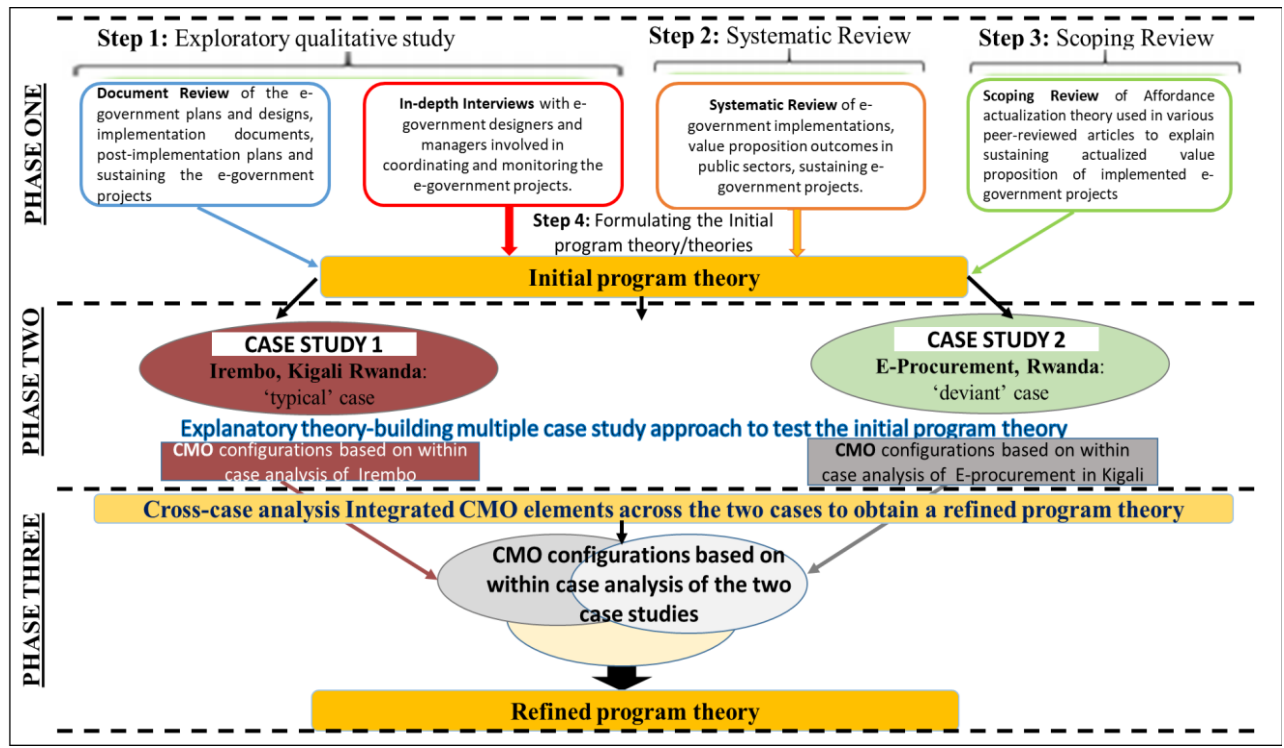

**Figure 1 Realist evaluation methodology as adapted from Mukumbang, Van Belle (2016)**

In Phase 1, we conducted a systematic literature review as well as documents review related to value propositions of implemented e-Government projects as noted earlier. Then gathered initial empirical evidence to support our initial program theory (IPT) development as shown in Figure 2. In doing this, we conducted face-to face interviews with managers of Irembo, and e-Procurement. We triangulated the interview data with (1) literature (2) participant's observation (3) archival document analysis (4) scoping review of affordance actualization guiding the understanding of the human-IT interaction and extractions of the value propositions outcomes seen as product in process (Layder, 1998). These data from the primary and secondary sources were used to develop the IPT. We anchored this process of IPT development through abductive data analysis that aligned with Layder (1998) version of adaptive theory development. Thus, we adopted the dialectical interplay where prior theoretical concepts of affordance actualization (Strong et al., 2014), Self-determination theory (SDT) (Deci & Ryan, 1985), Sustainability framework for e-Government success (Lessa et al., 2015) and context-mechanism-outcomes-configuration (Pawson et al., 2005) are integrated (Hoon & Baluch, 2019; Modell et al., 2017; Okhuysen & Bonardi, 2011; Ononiwu et al., 2018). Altogether, such integration of theories "shape and inform the analysis of data that emanate from the ongoing research at the same time the emergent data itself shapes and molds the existing theoretical materials" (Rambaree, 2018). Therefore, we adopted the abductive thematic analysis that involved four key steps of gathering the initial empirical evidence as explained earlier (Rambaree, 2018; Richter et al., 2018). Such steps which are recursive between collection and analyzing of data as well as being iterative until data saturation. Following Glaser and Strauss (1967) "saturation' means that no additional data are being found whereby the sociologist can develop properties of the category. As he sees similar instances over and over again, the researcher becomes empirically confident that a category is saturated." (p. 61).

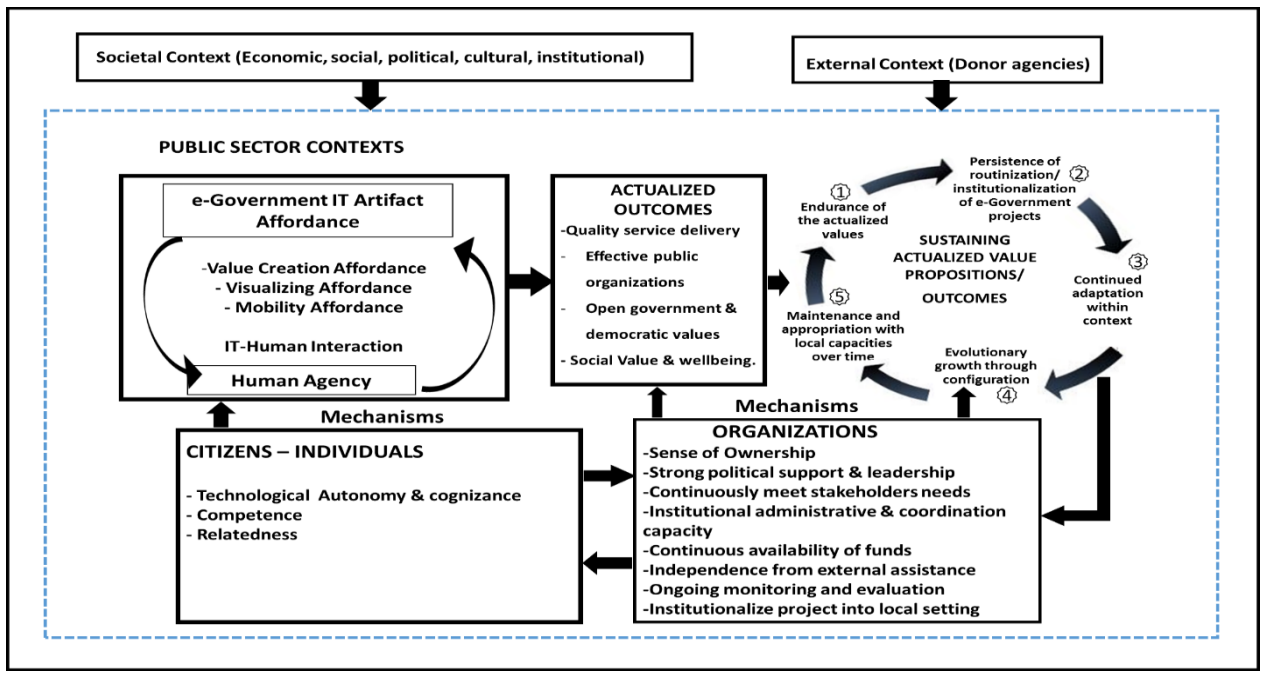

**Figure 2 Initial Program Theory (IPT) expounded from literature review, and initial data collected from multiple investigative contexts subject to refinement. Source: Authors**

## 4.1 Phase 2: Theory validation and refinement

We used both quantitative and qualitative methods to validate the IPT fusing the two investigative contexts as the multiple cases as shown in Figure 1. First, we conducted a quantitative survey using contingency evaluation method (CVM). This is to confirm if the citizens' actualized value propositions of implemented e-Government project exist. CVM survey instrument indicating citizens willing to pay to actualize or continue actualizing the value propositions of quality service delivery, effective public organizations, social value and general wellbeing and Open government and democratic values. The CVM instrument was developed to generate categorical data of Yes or NO when completed by the citizens. We distributed 430 questionnaires, with 250 of the questionnaires translated into Kinyarwanda, the local language of Rwanda and mostly used medium in engaging the e-Government platforms. Subsequently, we re-translated the answered questionnaires in Kinyarwanda back to English language. The remaining 150 questionnaires were in English and 30 are specifically for e-Procurement users, particularly, businesses that uses the platform. Subsequently, we conducted face to face and semi structured realist interviews. This is where the IPT are discussed with interviewees, audio recorded and transcribed verbatim. That lasted for about 30-45 minutes each with program designers, implementers, and managers of Irembo, and e-Procurement.

**Investigative Context**

Such case studies hold two dominant entities: typical and deviant. The "typical" case which is Irembo in Rwanda because it is an ideal one-stop platform that provide government-to-citizen (G2C), government-to-business (G2B), and government-to-government (G2G) services. The "deviant" case is the e-procurement platform in Rwanda. This is because it is an e-procurement platform accessed predominantly by businesses that are into supplies and contract biddings that witnessed initial resistance before being accepted.

## 4.3 Irembo

Irembo (meaning "gateway" in local parlance) is a public-private partnership (PPP) initiative conceived by the Government of Rwanda and Rwanda Online Platform Ltd (ROL). It is aimed at providing a single platform window for all government services (Bakunzibake et al., 2019). ROL will implement a digital one-stop platform for all government services across G2C, G2B, G2G, and to a large extent government-to-society (G2S) (Mukamurenzi et al., 2019). Services targeted were; registration such as birth, death, marriages, driver's license, and business registration. Other services include tax filing, land management, road traffic, and motor vehicle inspection (Twizeyimana, 2017). By September 28, 2017, 95 online services were fully deployed and accessible (Twizeyimana et al., 2018). Access to Irembo services is via mobile phones, USSD, computers or a visit to Irembo service centers, and Irembo agents.

### 4.4 e-Procurement

The government of Rwanda have changed its procurement processes from paper based to a digital e-Procurement platform called "Umucyo" meaning transparency (Harelimana, 2018). This is in line with global best practice in procurement used to adopt and retain the fundamental principles of transparency, accountability, competition, equity, and fairness (Omwono et al., 2020). The initiative was deployed in August 2016 with the inaugural system having eight public institutions. Such institutions include Rwanda public procurement authority (RPPA), Rwanda development board and ministry of finance and economic planning. Other institutions include ministry of justice, Rwanda revenue authority, Rwanda social security board, banks and insurance companies (Gihozo, 2020). Owing to its launching in 2016, e-Procurement (Umucyo) became the only avenue for all public procurement process in Rwanda by both public and private institutions (Isimbi U., 2016). Such as government purchases, goods, works, services, and non-consultancy to be integrated using Umucyo (Gihozo, 2020).

## 5. FINDINGS

We used logistic regression to analyze our data. It is a multivariate regression that permits the understanding of what independent variables could constitute and a dependent variable via their degree of importance, polarity of the effects and significant effects (Yilmaz & Belbag, 2016). Here the independent variables are quality service delivery, effective public organizations, social value and general wellbeing, and open government and democratic values. The dependent variable is the actualized value propositions seen as outcomes. Logistics regression is beneficial to the study for the interpretation of binary and categorical data (Suthar et al., 2010; Yilmaz & Belbag, 2016). Table 2 and 3 depict the result of the data analysis for Irembo and e-procurement in Rwandan e-Government implemented projects.

Based on the logistics regression, the **Wald** value should be ≥2 to show which variable have influence on the model. The value of **β** informs decision maker about the influence direction of the variable/factor. A negative value represents negative effect, while a positive value represents positive effect. **SIG** represents the significance value that should be less than 0.05 to be important in the model, while **EXP(β)** is the ranking order level of importance of the independent variables in the model. A variable with the highest **EXP(β)** value is the first most important variable in the model (Suthar et al., 2010; Yilmaz & Belbag, 2016).

| Independent Variables | Service | Item | β | S.E. | Wald | df | SIG. | Exp (β) |
|---|---|---|---|---|---|---|---|---|
| Quality Service Delivery | IS | Improved service delivery | 4. 235 | 1.654 | 17.269 | 1 | 0.25 | 74.986 |
| | IA | Improved access to services | 3.745 | .894 | 23.963 | 1 | .033 | 54.76 |

| | | | β | S.E. | Wald | df | SIG. | Exp (β) |
|---|---|---|---|---|---|---|---|---|
| | IT | Improved transparency, accuracy, and facilitating information between government and customers | 1.364 | .992 | 11.891 | 1 | .001 | 13.912 |
| | RC | Registration and certification | 1.358 | 1.071 | 1.608 | 1 | 0.35 | 13.887 |
| Effective Public Organizations | FT | Filing Tax | -17.669 | 2.911 | 14.000 | 1 | .097 | 1.000 |
| | NA | News and announcements | 8.192 | .373 | 15.0300 | 1 | .018 | 16.763 |
| | AT | Accountability and transparency | 6.640 | .486 | 18.050 | 1 | .027 | 56.070 |
| | SP | Streamlining procedures | -.622 | 1.263 | 16.000 | 1 | .087 | 5.37 |
| | PRO | Procurement | 4.337 | 1.684 | 11.230 | 1 | .039 | 44.89 |
| | HIDS | Horizontal integration (different functions and services) | -.071 | 2.852 | 11.000 | 1 | 1 | .932 |
| | IIKM | Integrated information knowledge management | -16.874 | 1.721 | 7.100 | 1 | .029 | 12.000 |
| Social Value and Wellbeing | SGM | Social group profile management | -1.361 | 2.962 | 8.000 | 1 | 1 | 6.532 |
| | STR | Status reports of government projects | .352 | 1.432 | 14.230 | 1 | .044 | 31.422 |
| | BT | Building trust between government and citizens | .352 | .968 | 16.133 | 1 | .016 | 14.22 |
| Open Government and Democratic Value | OD | Opendata | 3.227 | 1.570 | 14.226 | 1 | .040 | 24.195 |
| | IPEP | Improved public engagement and enhanced participation | -14.437 | 1.824 | 10.270 | 1 | .023 | 6.080 |
| | CV | Citizen's voice (Medium to reach out to government complain and feedback) | -16.770 | .962 | 12.056 | 1 | 1.00 | 2.132 |
| Constant | | | 35.139 | | | | .998 | 4.901E+58 |

**Table 2 Variables in equation Table for Irembo**

| Independent Variables | Service | Item | β | S.E. | Wald | df | SIG. | Exp (β) |
|---|---|---|---|---|---|---|---|---|

| Factor | Code | Description | B | S.E. | Wald | df | Sig. | Exp(β) |
|---|---|---|---|---|---|---|---|---|
| Quality Service Delivery | IS | Improved service delivery | 4.235 | 1.526 | 3.526 | 1 | 0.25 | 74.986 |
|  | IA | Improved access to services | 3.745 | 1.876 | 9.876 | 1 | .033 | 54.76 |
|  | IT | Improved transparency, accuracy, and facilitating information between government and customers | 1.364 | 1.031 | 7.031 | 1 | .001 | 13.912 |
|  | RC | Registration and certification | 1.358 | 1.978 | 2.978 | 1 | 0.35 | 13.887 |
| Effective Public Organizations | FT | Filing Tax | -17.669 | .526 | 3.030 | 1 | .097 | 1.000 |
|  | NA | News and announcements | 8.192 | 3.005 | 13.005 | 1 | .018 | 16.763 |
|  | SP | Streamlining procedures | -.622 | 1.832 | 14.727 | 1 | .087 | 5.37 |
|  | PRO | Procurement | 4.337 | .237 | 19.399 | 1 | .039 | 44.89 |
| Social Value and Wellbeing | SGM | Social group profile management | -1.361 | .876 | 19.876 | 1 | 1 | 6.532 |
|  | STR | Status reports of government projects | .352 | .373 | 16.423 | 1 | .044 | 31.422 |
|  | BT | Building trust between government and citizens | .352 | 1.875 | 11.875 | 1 | .016 | 14.22 |
| Open Government and Democratic Value | OD | Opendata | 3.227 | 1.597 | 12.597 | 1 | .040 | 24.195 |
|  | CV | Citizen's voice (Medium to reach out to government complain and feedback) | -16.770 | 2.893 | 18.47 | 1 | 1.00 | 2.132 |
| Constant |  |  | 42.539 |  |  |  | .999 | 50.86 |

**Table 3 Variables in equation Table for e-Procurement**

As both two Tables 2 and 3 could attest, Quality Service Delivery, Effective Public Organizations, Social Value and General Wellbeing, and Open Government and Democratic values constitute the actualized value propositions. The factors have the **Wald** value ≥2 to show that such factors have influence on the model. With **SIG** less than 0.05 with most of the items that represent the factors, suggest that such factors influence the actualized values. While the higher **EXP(β)** suggests that all the factors constitute the actualized value propositions with item improved service delivery under quality of service as the highest.

### 5.1 Goodness-of-Fit of the Model

Goodness of Fit for both Irembo and eProcurement models is the measure of Likelihood. Likelihood is the probability of observed results; given the parameter estimates and -2 times the log likelihood (-2LL). The -2log likelihood is in turn the measure of badness-of-fit, illustrating error remaining in the model after accounting for all independent variables. The -2LL of 44.817 in the Irembo model indicated that there is no significant error remaining in the model. The Nagelkerke R square shows that about 80% of the variation in the actualized values is explained by Quality Service Delivery, Effective Public Organizations, Social Value and General Wellbeing, and Open Government and Democratic values as shown in Table 4.

| Step | -2 Log likelihood | Cox & Snell R Square | Nagelkerke R Square |
|---|---|---|---|
| 1 | 44.817a | .043 | .795 |

**Table 4 Model summary table for Irembo**

The same goes for eProcurement model, where the -2LL of 17.173 indicates that there is no significant error remaining in the model. The Nagelkerke R square shows that about 75% of the variation in the actualized values is explained by Quality Service Delivery, Effective Public Organizations. Social Value and General Wellbeing as well as Open Government and Democratic values as shown in Table 5.

| Step | -2 Log likelihood | Cox & Snell R Square | Nagelkerke R Square |
|---|---|---|---|
| 1 | 17.173a | .388 | .754 |

**Table 5 Model Summary for e-Procurement**

Having now empirically identified the factors that constituted the actualized values propositions in the investigative context, we move to how it was sustained as shown in the next section.

**5.2 Sustaining the Actualized Value Propositions**

Our empirical qualitative data analysis as shown in Table 6 confirmed that: (a) Endurance of the actualized values. (b) Persistence of routinization or institutionalization of the initial e-Government projects. (c) Continued adaptation within the context the innovation is embedded. (d) Evolutionary growth through configuration, and (e) Maintenance and appropriation with local capacities over time are factors that reflect sustaining the actualized values. Such factors identified empirically as what constitute sustainability of the actualized values.

| Factors | Direct Quotes from Participants |
|---|---|
| Endurance of actualized values | *"We continue to use the platform components and activities after the initial funding period for achievement of our intended outcomes"*  *Thematic Analysis from Realist interview with Irembo* |
| Persistence of routinization or institutionalization of the initial e-Government projects | *"We put in place structures and processes of innovation as a culture of practices by staff, the organizations and our systems"*  *Thematic Analysis from Realist interview with SORMAS* |
| Continued adaptation within the context the innovation | *"Our platform adapts to distinct possible situation within context of the services we deliver"*  *Thematic Analysis from Realist interview with SORMAS* |
| Evolutionary growth through configuration | *"We have in place organizational strategies, structures, and processes to grow and sustain the platform"*  *Thematic Analysis from Realist interview with e-Procurement* |
| Maintenance and appropriation with local capacities over time | *"We rebuilt and launched the platform to be fully managed by local staff"* Extract from Interview in Irembo |

**Table 6. Sustaining the Actualized Value Propositions**

Besides, since the actualized value propositions are the concrete outcome of acting on the affordance of the implemented e-Government projects. It then behooves us to identify such affordances that are useful for realizing the actualized value propositions that is under sustenance as detailed next.

### 5.3 Affordances of the implemented e-Government Projects

e-Government affordances as revealed by the data were value creation affordance (Manoury et al., 2019), visualizing affordance (Rietveld & Brouwers, 2017), as well as mobility affordance (Kohut, 2018) are empirically validated as shown in Table 5.

| Affordance Types | Direct Quotes from Participants |
|---|---|
| Value Creation Affordance | *"We are creating value through implementation of e-Procurement platform by simply not only automating the process with the value that automation brings but also to solve existing issues in public procurement" "We are increasing statistics, integrating open contracting for better standards and analytics, artificial intelligence" "This is increasing participation and participation increase competition, while competition increase transparency"* Source: Thematic Analysis from Realist interview with e-Procurement<br><br>*"So, I will say that the reason for implementing Irembo is to give citizens a better experience and the government to enhance value creation"* Source: Thematic Analysis from Realist interview with Irembo |
| Visualizing Affordance | *"The product team is the team in charge of understanding the users, and designing the product, the experience that will match the user's requirement that incorporate their peculiarities. Such peculiarities in design captures the visual function in using the platform. Irrespective of individual ability, the visual design enhances ease of use"* Source: Thematic Analysis from Realist interview with Irembo |
| Mobility Affordance | *"So, there's a web version and then the mobile version. So, for the mobile version you can go on without internet so they can list their cases on SORMAS when they go back to where there is internet then they synchronize. For the laptop version, obviously, you can choose it without internet"* Source: Thematic Analysis from Realist interview with SORMAS<br><br>*"Simply giving one single window for public procurement tenders with the information available from anywhere and to everyone at the same time"* Source: Thematic Analysis from Realist interview with e-Procurement |

**Table 7 e-Government Affordances**

Now under what context and mechanisms in charge of sustaining the actualized value propositions as outcomes is detailed next.

### 5.4 Context-Mechanism-Outcomes

As Table 6 could attest, our empirical data confirm the mechanisms are of two types. Individual mechanisms and organizational mechanism. The individual mechanisms are: (1) Technological autonomy and cognizance, and (2) Competence. The organizational mechanisms are: (1) Continuous availability of funds, (2) Institutional administrative and coordinating capacity, (3) Sense of ownership, (4) Strong political support and leadership, (5) Institutionalized in local settings, and (6)

Ongoing monitoring and evaluation. The contextual conditions are socio-economic, institutional, political and external donor agencies.

| Context | Mechanism | Outcomes validated by the logistic regression analysis of CVM data | Outcome Category |
|---|---|---|---|
| Societal<br><br>-Economic<br><br>*With the "deployment of the platform, it becomes the central payment collection point for all government services" Source: Thematic Analysis of Realist interview* | Continuous availability of funds<br><br>*"We have a business model that charges commission on every transaction" Source: Thematic Analysis of Realist interview* | Improved service delivery | Quality Service Delivery |
| -Social<br><br>*"Irembo is a One-stop platform for all government services" Source: Thematic Analysis of Realist interview* | Institutional administrative and coordinating capacity<br><br>*"We are headed by a DG, directors of product services, engineering, developers, and support team thereby institutionalizing our capacity to run the platform" Source: Thematic Analysis of Realist interview*<br><br>Sense of ownership *"We have multiple service channels (website, USSD, agents) in English, Kinyarwanda, and French making the platform to have a sense of local ownership" Source: Thematic Analysis of Realist interview* | Improved service delivery<br><br>Opendata<br><br><br>Social group profile management<br><br>Building trust between government and citizens<br><br>Status reports of contact tracing, isolation, and quarantine<br><br>Status reports of surveillance and response<br><br>Status reports on diseases and pandemic | Quality Service Delivery<br><br>Open Government and Democratic Value<br><br>Social Value and General Wellbeing<br><br>Social Value and General Wellbeing<br><br>Social Value and General Wellbeing<br><br>Social Value and General Wellbeing<br><br>Social Value and General Wellbeing |
| Political<br><br>*"Part of a larger National ICT roadmap and Master Plan being supported by political leadership of council of ministers" Thematic Analysis of Realist interview* | Strong political support and leadership *"We have a dedicated Minister of ICT who is responsible for the success of the platform" Source: Thematic Analysis of Realist interview* | | |
| -Institutional<br><br>"The Rwanda Public Procurement Law was revised in August 2018 (No 62/2018) to incorporate the e-procurement system"<br><br>*Source: Thematic Analysis from Realist interview* | Institutionalized in local settings<br><br>"We rebuilt and launched the platform to be fully managed by local staff"<br><br>*Source: Thematic Analysis of Realist interview* | Improved transparency, accuracy, and facilitating information between government and citizens<br><br>Accountability and transparency<br><br>Streamlining procedures<br><br>Procurement | Effective Public Organizations<br><br>Effective Public Organizations<br><br>Effective Public Organizations<br><br>Effective Public Organizations |
| External-Donor agencies<br><br>"The platform was sponsored by World Bank"<br><br>*Source: Thematic Analysis from Realist interview* | Ongoing monitoring and evaluation<br><br>"We have a support team for feedback through social media platforms, email, and toll free call centers to ensure monitoring and evaluation"<br><br>*Source: Thematic Analysis of Realist interview* | Improved transparency, accuracy, and facilitating information between government and citizens<br><br>*Source: Data from Contingent Valuation Questionnaire* | Effective Public Organizations |
| | Technological autonomy and cognizance<br>"Knowing that the platform is a secured automated procurement process gives me the confidence and convenience to select which tender to bid for anywhere, anytime" | *Source: Data from Contingent Valuation Questionnaire on rationale for choice under actualize value and willingness to pay* | |
| | Competence<br>"My IT skills and ability enable easy use of the platform which in turn save time, resources, and to get update on news and information" | *Data from Contingent Valuation Questionnaire on rationale for choice under actualize value and willingness to pay* | |

**Table 8 Context-Mechanism-Outcome configuration Table.** *Source:* **Authors**

## 5.5 Phase 3: Theory Consolidation

In our third phase, we synthesized the findings into one model as shown in Figure 3.

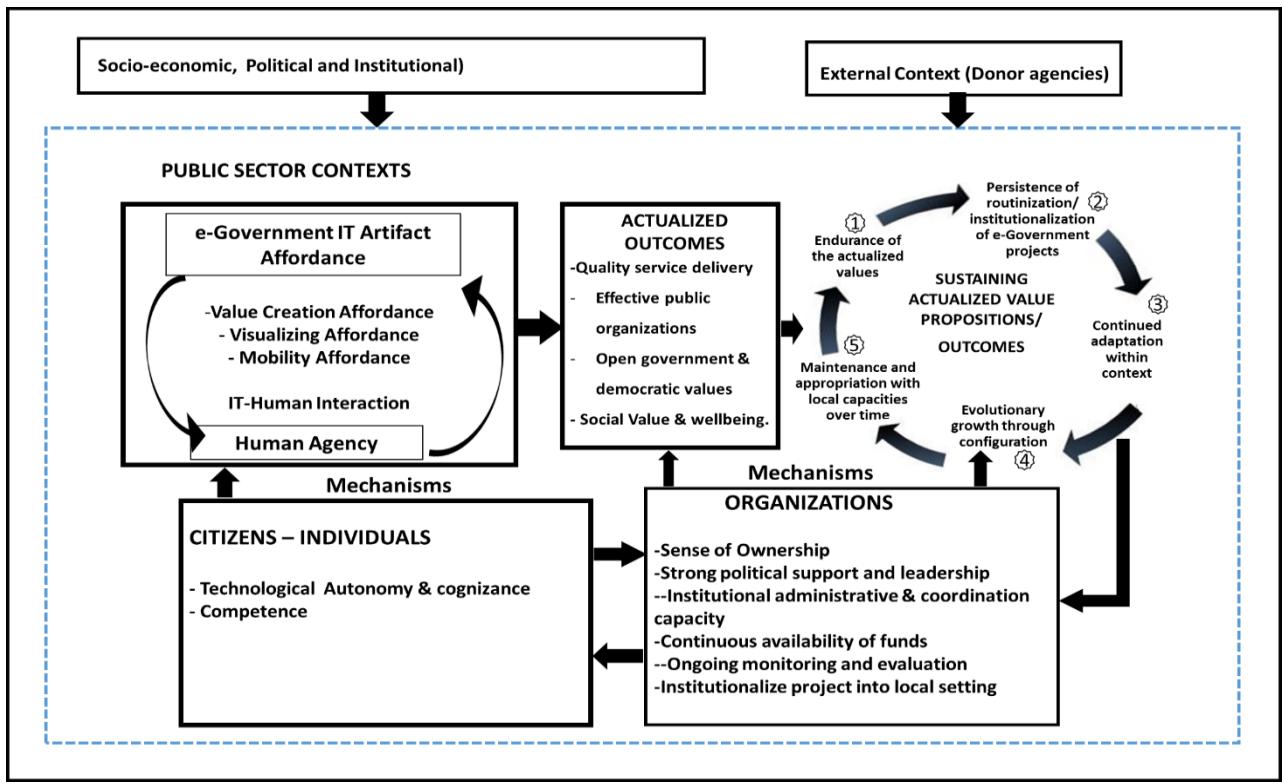

**Figure 3: Refined Model of Actualized Values Sustainability Framework of Implemented e-Government Projects.  Source: Authors**

The mechanisms seen in the IPT of Figure 2 such as independence from external assistance, continuously meeting the stakeholders' needs and relatedness were eliminated due to insufficient empirical data to support them.

## 6. DISCUSSION

Based on our findings, the actualized value propositions of the implemented e-Government projects in the investigative context is constituted with the following factors: (a) Quality service delivery, which is also confirmed by other scholars  (e.g., Bannister & Connolly, 2014; Castelnovo, 2013). (b) Effective public organization (Heeks & Bailur, 2007). (c) Open government and democratic values (Deng et al., 2018; Harrison et al., 2012), and lastly (d) Social value and general wellbeing (Grimsley & Meehan, 2008; Karunasena & Deng, 2012). Such factors as outcomes came to be by affordance actualization of the e-Government projects. The affordances that have been actualized include value creation affordance, visualizing affordance and mobility affordance. This is done by the citizens' competences in the technology as one of the individual mechanisms that must be in operation. Such citizens are also endowed with technological autonomy and cognizance in the e-Government environment when engaging with the technology as another individual mechanism. Besides, the organization (i.e., the public sector that host the e-government projects) is endured with continuous availability of funds by the host government, institutional administrative and coordinating capacity with strong leadership and political support from the government in power as mechanisms. Other mechanisms were the technology being institutionalized in the local settings with its ongoing monitoring and evaluation for improvements. Such identified underlying mechanisms are in concerted interaction to actualize the e-Governments affordances to bring about the actualized value propositions as outcomes and its sustainability. Empirically, the sustainability of the actualized values takes the shape of endurance of the identified actualized values, persistence of institutionalization of the e-Government projects and continued adaptation within the context the innovation is embedded. Other of such shapes of sustainability include evolutionary growth through configuration, as well as maintenance and appropriation with local capacities over time. All of these occurred when external donor agencies such as the World Bank continue to support the host county as the enabling condition. Other enabling conditions include enactment of institutional policies, political and socio-economic enablement. All these are encapsulated in the consolidated model shown in Figure 3 and classified as the actualized values sustainability framework of implemented e-Government projects.

## 7. CONTRIBUTIONS

By the use of realist evaluation in e-Government studies, our nascent refined program theory in the form actualized values sustainability framework of implemented e-Government projects could provide the following contributions. The framework could be generalized to other emerging economies of sub-Saharan Africa to be used for identification of actualized values propositions of implemented e-Government projects their citizens are willing to pay (WTP) in their contexts-of-use. By discovering a new research stream and unexplored opportunities through the development of a new research framework (i.e., actualized values sustainability framework), we open the door for future investigations in e-Government research. By the use of such a framework, researchers will identify e-Government affordances and both organization and citizens' motivations and capabilities as mechanisms to actualize such affordances in order to obtain the values as outcomes and sustaining it. To practitioners it will: (1) build-up citizens' satisfaction and public trust about government e-service delivery (Agbabiaka, 2018), (2) meet the demands for external/donor accountability and (3) establish a clear, strategic goal for the organization. Other benefits to practitioners are: (4) identify the relevant value propositions from the citizen's perspective to foster a strong sense of operational accountability (Agbabiaka, 2018; Panagiotopoulos et al., 2019), and lastly (5) to increase organizational performance (Panagiotopoulos et al., 2019).

## 8. CONCLUSSION AND FUTURE RESEARCH

Based on this study, we come to know that actualization of the value proposition of implemented e-Government projects and sustaining it is critical in Sub-Saharan Africa. This is because citizens have to engage and use such projects to achieve their needs in their respective contexts with satisfaction if they are to trust the government. Besides, other stakeholders such as businesses must continuously use the e-Government projects to achieve the cardinal objectives of their implementations. Consequently, by developing the actualized values sustainability framework of implemented e-Government projects we have unraveled what is really the actualized values and citizens' motivation to use the platforms (Deci & Ryan, 1985; Ryan & Deci, 2002). Besides, we revealed the generative mechanisms to sustain the value propositions to shape innovation continuation. As we deployed the realist evaluation methodology Mukumbang et al. (2016) that resonated with multiple theories; we come to understand the enabling contexts, the concerted mechanisms in interaction both at individual and organizational levels to sustain the actualized value propositions as outcomes. The future research stream is to adopt the actualized values sustainability framework in other countries of sub-Saharan Africa that have implemented e-Government projects.